# Combined Integer and Floating Point Multiplication Architecture(CIFM) for FPGAs and Its Reversible Logic Implementation

Himanshu Thapliyal
Centre for VLSI Design
IIIT Hyderabad, India
(thapliyalhimanshu@yahoo.com)

Hamid R. Arabnia
The University of Georgia,
Department of Computer Science, USA
(hra@cs.uga.edu)

A.P Vinod
School of Computer Engineering
Nanyang Technological University,
Singapore (asvinod@ntu.edu.sg)

*Abstract*— **In this paper, the authors propose the idea of a combined integer and floating point multiplier(CIFM) for FPGAs. The authors propose the replacement of existing 18x18 dedicated multipliers in FPGAs with dedicated 24x24 multipliers designed with small 4x4 bit multipliers. It is also proposed that for every dedicated 24x24 bit multiplier block designed with 4x4 bit multipliers, four redundant 4x4 multiplier should be provided to enforce the feature of self repairability (to recover from the faults). In the proposed CIFM reconfigurability at run time is also provided resulting in low power. The major source of motivation for providing the dedicated 24x24 bit multiplier stems from the fact that single precision floating point multiplier requires 24x24 bit integer multiplier for mantissa multiplication. A reconfigurable, self-repairable 24x24 bit multiplier (implemented with 4x4 bit multiply modules) will ideally suit this purpose, making FPGAs more suitable for integer as well floating point operations. A dedicated 4x4 bit multiplier is also proposed in this paper. Moreover, in the recent years, reversible logic has emerged as a promising technology having its applications in low power CMOS, quantum computing, nanotechnology, and optical computing. It is not possible to realize quantum computing without reversible logic. Thus, this paper also paper provides the reversible logic implementation of the proposed CIFM. The reversible CIFM designed and proposed here will form the basis of the completely reversible FPGAs.**

## I. INTRODUCTION

Image and digital signal processing applications require high floating point calculations throughput, and nowadays FPGAs are being used for performing these Digital Signal Processing (DSP) operations. Floating point operations are hard to implement on FPGAs as their algorithms are quite complex [1]. In order to combat this performance bottleneck, FPGAs vendors including Xilinx have introduced FPGAs with nearly 254 18x18 bit dedicated multipliers [2]. These architectures can cater the need of high speed integer operations but are not suitable for performing floating point operations especially multiplication. Floating point multiplication is one of the performance bottlenecks in high speed and low power image and digital signal processing applications [3]. Recently, there has been significant work on analysis of high-performance floating-point arithmetic on FPGAs[7,8,9,10]. But so far no one has addressed the issue of changing the dedicated 18x18 multipliers in FPGAs by an alternative implementation for improvement in floating point efficiency. It is a well known concept that the single precision floating point multiplication algorithm is divided into three main parts corresponding to the three parts of the single precision format. In FPGAs, the bottleneck of any single precision floating-point design is the 24x24 bit integer multiplier required for multiplication of the mantissas. In order to circumvent the aforesaid problems, this paper proposes a novel combined integer and floating point multiplication architecture (CIFM). The CIFM can perform both integer as well as single precision floating point multiplication with a single dedicated 24x24 bit multiplier block designed with small 4x4 bit multipliers. The basic idea is to replace the existing 18x18 multipliers in FPGAs by dedicated 24x24 bit multiplier blocks which are implemented with dedicated 4x4 bit multipliers, making the FPGAs suitable for integer as well as floating point calculations. The proposed architecture also brings the idea of reconfigurability and self repairability [6] at runtime, thus providing a low power as well as fault recovering architecture in FPGAs. The proposed architecture is especially designed for high performance and low power floating point multiplications in FPGAs. Since, the authors propose the idea of implementing CIFM with dedicated 24x24 bit multiplier designed with small 4x4 bit multipliers. Hence, a novel dedicated 4x4 bit multiplier beneficial in terms of speed, power and area is also proposed in this paper. Furthermore, researchers like Landauer have shown that for irreversible logic computations, each bit of information lost, generates $kT\ln 2$ joules of heat energy, where k is Boltzmann's constant and T the absolute temperature at which computation is performed [11]. Bennett showed that $kT\ln 2$ energy dissipation would not occur, if a computation is carried out in a reversible way [12], since the amount of energy dissipated in a system bears a direct relationship to the number of bits erased during computation. Reversible circuits are those circuits that do not lose information and reversible computation in a system can be performed only when the system comprises of reversible gates. These circuits can generate unique output vector from each input vector, and vice versa, that is there is a one-to-one mapping between input and output vectors. Thus, an NXN reversible gate can be represented as Iv=(I1, I2,I3,I4,…………………….……………………IN).
Ov = (O1, O2 , O3,…………………...………………..ON).

Where Iv and Ov represent the input and output vectors respectively. Classical logic gates are irreversible since input vector states cannot be uniquely reconstructed from the output vector states. There are a number of existing reversible gates such as Fredkin gate [13], TSG [4,5] and the New Gate (NG) [14]. As the Moore's law continues to hold, the processing power doubles every 18 months. The current irreversible technologies will dissipate a lot of heat and can reduce the life of the circuit. The reversible logic operations do not erase (lose) information and dissipate very less heat. Thus, reversible logic is likely to be in demand in high speed power aware circuits. Reversible circuits are of high interest in low-power CMOS design, optical computing, nanotechnology and quantum computing. It has been proved that the quantum arithmetic must be built from reversible logical components. The major constraints in reversible logic are

1. to minimize the number of reversible gates.
2. to minimize the number of garbage outputs. (Garbage output refers to the output that is not used for further computations).

This paper also introduces the reversible logic implementation of the proposed CIFM using a recently proposed TSG gate [4,5] and New gate[14]. The TSG gate has the advantage that it can work singly as a reversible Full adder with only two garbage outputs while the New Gate has the advantage that it can work singly as reversible half adder with bare minimum of one garbage output. Thus the highly optimized reversible implementation of the CIFM is proposed, best in terms of number of reversible gates and garbage outputs. It can be considered as an attempt to provide a primitive prototype of components of reversible FPGAs.

## II. FLOATING POINT MULTIPLIER ARCHITECTURE

The single precision floating point algorithm is divided into three main parts corresponding to the three parts of the single precision format. The first part of the product which is the sign is determined by an exclusive OR function of the two input signs. The exponent of the product which is the second part is calculated by adding the two input exponents. The third part which is the significand of the product is determined by multiplying the two input significands each with a "1" concatenated to it. Figure 1 shows the architecture of the single precision floating point multiplier. It can be easily observed from the Figure 1 that 24x24 bit integer multiplier is the main performance bottleneck for high speed and low power operations. In FPGAs, the availability of the dedicated 18x18 multipliers instead of dedicated 24x24 bit multiply blocks further complicates this problem. This is the driving force that leads to the proposal of CIFM architecture suitable both for integer as well as floating point multiplication operations.

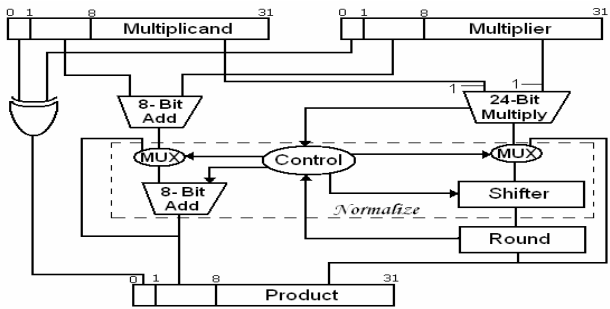

Figure 1. Single Precision Floationg Point Multiplication Architecture

## III. PROPOSED CIFM ARCHITECTURE FOR MULTIPLICATION IN FPGAS

The authors propose the idea of a combined integer and floating point multiplier (CIFM) for FPGAs. In the CIFM, it is proposed to replace the existing 18x18 bit multipliers in FPGAs with dedicated blocks of 24x24 bit integer multipliers designed with 4x4 bit multipliers(CIFM). The reason for this stems from the fact that it will make the FPGAs also suitable for floating point multiplication operations. The features of reconfigurabilty and self repairability at run time are also proposed in the architecture to attain low power and self repairability from faults.

### A. Reconfigurability Feature

In the proposed architecture, the dedicated 24x24 bit multiplication block is fragmented to four parallel 12x12 bit multiplication modules as shown in Figure 2, where AH, AL, BH and BL are each of 12 bits. The 12x12 multiplication modules are implemented using small 4x4 bit multipliers as shown in Figure 3. Thus, the whole 24x24 bit multiplication operation is divided into 36 4x4 multiply modules working in parallel. As shown in Figure 2, the proposed 24x24 bit multiplication architecture is reconfigurable at run time with the outputs of checkers working as control signals. If any of (A or B)'s mantissa is only of 12 bits then the Checker will check this and will switch of the multiply blocks which are not required using the control signal. Thus significant power saving can be attained at run time (on fly). The reconfigurability at run time for attaining low power has also been extended to individual 12x12 bit multiply modules.

As shown in Figure 3, the 12 bit numbers A & B to be multiplied are divided into 4 bits groups A3,A2,A1 and B3,B2,B1 respectively. Checkers at A3,A2 and B3,B2 will check whether the mantissas to be multiplied are of 12 bits, 8 bits or 4 bits. Then accordingly, will switch on or switch off, the required 4x4 multiply modules. Hence, there is a significant reduction in power consumption if the numbers to be multiplied are less than 12 bits, as only the required blocks are operating while others are switched off.

### B. Self Repairability

Self repairability at run time is also provided by providing a redundant 4x4 multiply module to each 12x12 multiply module, as shown in Figure 4. The product of the redundant multiplier is distributed to all 4x4 bit multiplier blocks making the 12x12 bit multiply module. The 4x4 multiplier to be repaired is specified by the given Aij, Bij and E bits. Then the 4x4 multiplier to be repaired abandons its own output and replaces it by the one from the extra multiplier. It should be noticed that the power supply of the disabled unit (one of the nine 4x4 multiplier) will be turned off through a power enable control to reduce the power dissipation. Thus, the proposed multiplier is also capable of recovering from faults.

### C. Additional Advantages

The additional advantage of the proposed CIFM is that floating point multiplication operation can now be performed easily in FPGA without any resource and performance bottleneck. In the single precision floating point multiplication, the mantissas are of 23 bits. Thus, 24x24 bit (23 bit mantissa +1 hidden bit) multiply operation is required for getting the intermediate product. With the proposed architecture, the 24x24 bit mantissa multiplication can now be easily performed by passing it to the dedicated 24x24 bit multiply block, which will generate the product with its dedicated small 4x4 bit multipliers. If either of the mantissa are less than 23 bits, reconfigurability feature at run time in the proposed CIFM will help in achieving significant power saving. Moreover, the redundant multipliers in the 24x24 bit multiply block will also take care of fault in any of the dedicated 4x4 bit multipliers. The large integer multiply operations can also be performed easily by dedicated 4x4 bit multipliers reducing the need of dedicated large size multipliers.

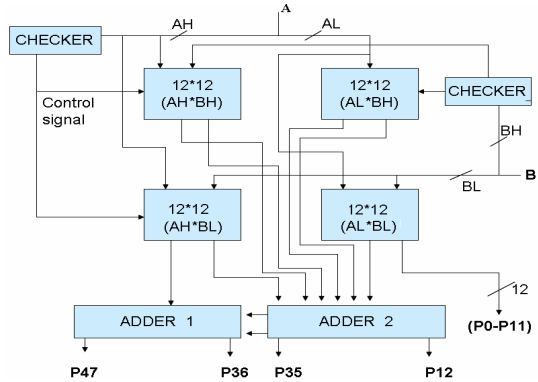

Figure 2. Proposed 24x24 bit Architecture

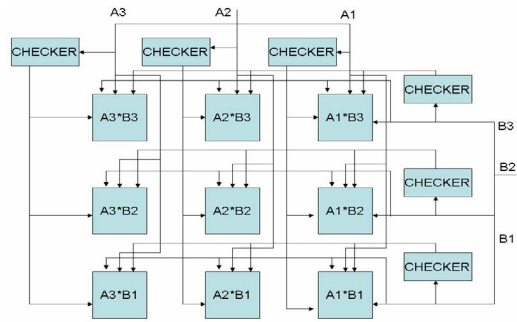

Figure 3. Internal structure of Individual 12x12 multiply module

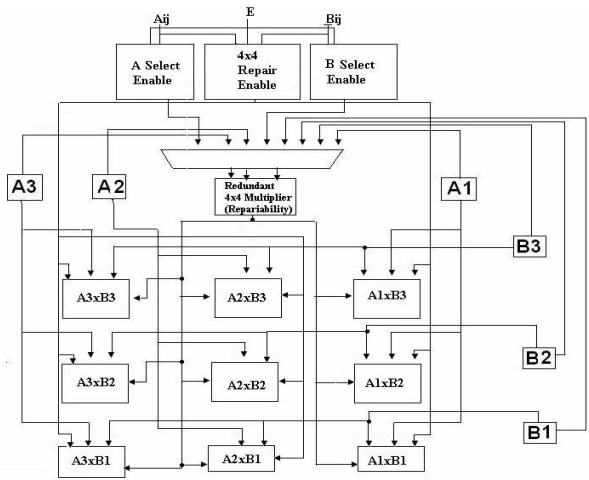

Figure 4. Proposed Feature of Self Repairability

## IV. PROPOSED DEDICATED 4X4 BIT MULTIPLIER

As evident from the proposed CIFM architecture, a high speed low power dedicated 4x4 bit multiplier will significantly improve the efficiency of the CIFM architecture. Thus, a dedicated 4x4 bit multiplier efficient in terms of area, speed and power is proposed. Figure 5 shows the architecture of the proposed multiplier. For (4 X 4) bits, 4 partial products are generated, and are added in parallel. Each two adjacent partial product are subdivided to 2 bit blocks, where a 2 bit sum is generated by employing a 2-bit parallel adder appropriately designed by choosing the combination of half adder-half adder, half adder-full adder( forming the blocks 1,2,3,4 working in parallel). This forms the first level of computation. The partial sums thus generated are added again in block 5 & 6 (parallel adders), working in parallel by appropriately choosing the combination of half adders and full adders. This forms the second level of computation. The partial sums generated in the second level are utilized in the third level(blocks 7 &8) to arrive at the final product. Hence, there is a significant reduction in the power consumption since the whole computation has been hierarchically divided to levels. The reason for this stems from the fact that power is provided only to the level that is involved in computation and thereby rendering the remaining two levels switched off (by employing a control circuitry). Working in parallel significantly improves the speed of the proposed multiplier. The proposed architecture is highly optimized in terms of area, speed and power. The proposed architecture is functionally verified in Verilog HDL and synthesized in Xilinx FPGA.

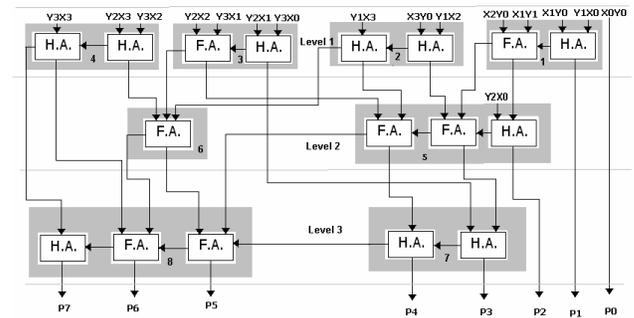

Figure 5. Proposed Dedicated 4x4 Bit Multiplier

## V. VERIFICATION AND IMPLEMENTATION

In this study, the proposed CIFM architecture is implemented in Verilog HDL and logic simulation is done in Veriwell Simulator; the synthesis and FPGA implementation is done using Xilinx Webpack 6.1. The design is optimized for speed and area using Xilinx , Device Family : VirtexE, Device : XCV300e, Package: bg432, Speed grade: -8. The device is made up of multiplexers and LUTs. FPGA synthesis results have shown that the proposed feature of reconfigurability at run time and the control circuitry designed for the introduction of this feature will marginally increase the delay and area of the 24x24 bit dedicated block. It has been found that for Xilinx VirtexE family, the delay of the proposed architecture is 41.203 ns with the area(cell usage) of 3149 while the delay is 37.553 ns with the area(cell usage) of 2967 without the additional feature of repairability. The results are shown in Table 1. Thus the results show that there is an increment of 9.71% in delay and 6.13% in area with the introduction of feature of reconfigurability at run time which can be considered negligible with the advantages associated with it.

TABLE I. SYNTHESIS RESULTS OF THE PROPOSED CIFM ARCHITECTURE

| Name of Multiplier | Vendor | Device Family & Device | Package | Speed Grade | Cell Use | Estimatd Delay (ns) |
|---|---|---|---|---|---|---|
| Proposed Multiplier Without Reconfigurability | Xilinx | VirtexE Xcv300e | Bg432 | -8 | 2967 | 37.553 |
| Proposed Multiplier With Reconfigurability | Xilinx | VirtexE Xcv300e | Bg432 | -8 | 3149 | 41.203 |

## VI. REVERSIBLE LOGIC IMPLEMENTATION OF PROPOSED CIFM

In order to implement the reversible logic design of the proposed CIFM, some of the basic concepts of the reversible logic are discussed.

### A. Basic Reversible Gates

There are a number of existing reversible gates such as Fredkin gate [13], TSG [4,5] and the New Gate (NG) [14]. Since, the major reversible gate used in designing the reversible CIFM is TSG gate, hence only the TSG gate is discussed in this section.

#### 1) TSG GATE

Recently, a 4 * 4 one through reversible gate called TS gate "TSG" is proposed [4,5,6]. The reversible TSG gate is shown in Fig. 6. The TSG gate can implement all Boolean functions.

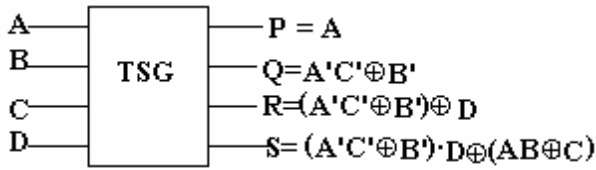

Figure 6. Reversible 4 *4 TSG proposed in [4,5,6]

One of the prominent functionality of the TSG gate is that it can work singly as a reversible Full adder unit. Fig. 7 shows the implementation of the TSG gate as a reversible Full adder.

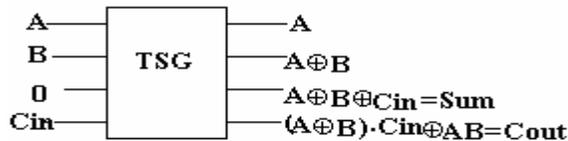

Figure 7. TSG Gate working Singly As a Reversuble Full Adder

A number of reversible full adders were proposed in [14,15,16,17]. The reversible full adder circuit in [14] requires three reversible gates (two 3*3 new gate and one 2*2 Feynman gate) and produces three garbage outputs. The reversible full adder circuit in [15,16] requires three reversible gates (one 3*3 new gate, one 3*3 Toffoli gate and one 2*2 Feynman gate) and produces two garbage outputs. The design in [17] requires five reversible Fredkin gate and produces five garbage outputs. The full adder designed using TSG in Fig. 5 requires only one reversible gate (one TSG gate) and produces only two garbage outputs. Hence, the full-adder design in Fig. 5 using TSG gate is better than the previous full-adder designs of [14,15,16,17]. A comparative experimental result is shown in Table II.

TABLE II. EXPERIMENTAL RESULTS OF DIFFERENT REVERSIBLE FULL ADDER CIRCUITS

|  | No of Gates | No of Garbage Outputs | Unit Delay |
|---|---|---|---|
| Full adder Using TSG | 1 | 2 | 1 |
| Existing Circuit[14] | 3 | 3 | 3 |
| Existing Circuit [15,16] | 3 | 2 | 3 |
| Existing Circuit[17] | 5 | 5 | 5 |

### B. Design of Reversible CIFM

As evident from the architecture of CIFM, the primary requirements to design reversible CIFM are the reversible 4x4 bit multiplier and reversible parallel adders. The authors have already designed high speed optimal reversible parallel adders highly optimized in terms of number of reversible gates and garbage outputs [4,5,6]. For the reversible implementation of the proposed 4x4 bit dedicated multiplier, reversible full adder and half adder are required. As shown above TSG gate is the best gate to design reversible full adder but if we use TSG gate to design reversible half adder, the garbage outputs will increase in the proposed design. Thus, we have used New Gate [14] to design reversible half adder to make the design highly optimized in terms of number of reversible gates and garbage outputs. The New gate can realize the half adder with bare minimum of one garbage output. Figure 8 (a) shows the New Gate and Figure 8(b) shows its working as a reversible half adder.

Figure 9 shows the reversible implementation of the proposed 4x4 dedicated multipliers. Once you have the reversible 4x4 dedicated multipliers, the reversible 12x12 multipliers can be easily designed from them as explained earlier to finally generate the CIFM architecture as shown in Figure 10.

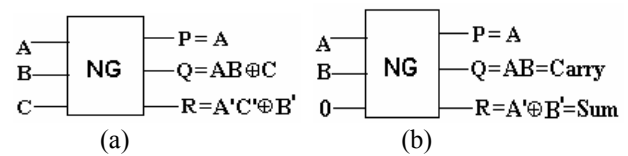

(a)                                         (b)

Figure 8. (a) New Gate(NG) (b) New Gate as Half Adder

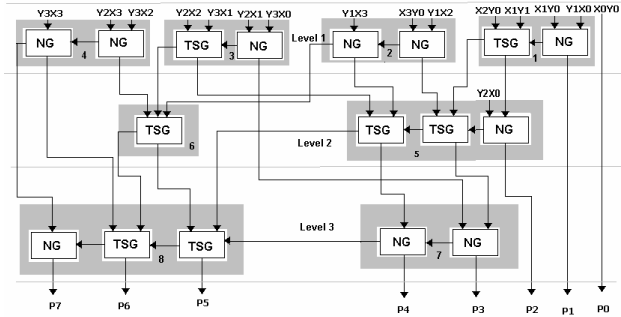

Figure 9. Reversible implementation of proposed dedicated 4x4 bit Multiplier( Garbage outputs not shown )

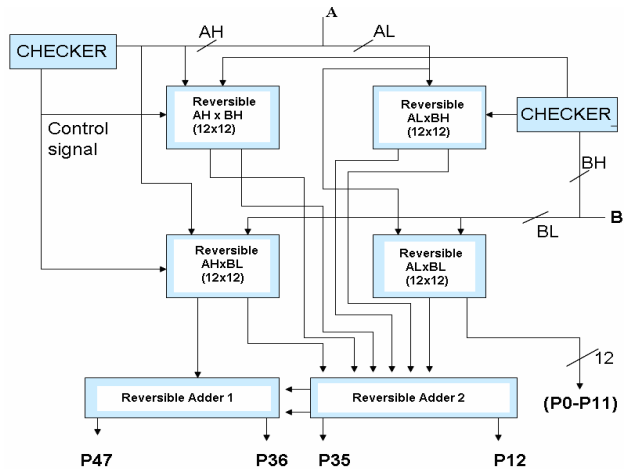

Figure 10. Proposed Reversible CIFM Architecture

## VII. CONCLUSION

This paper proposes a highly regular self-repairable and reconfigurable at run time(on fly) combined integer and floating point multiplication architecture (CIFM) for performing both integer as well as single precision floating point multiplication. Both the repairing and reconfigurability take the advantage of the partitioning of the circuit, which results in high controllability and observability, inherent in the decomposition approach. The results obtained are quite encouraging and there is a marginally increase in area and the delay of the CIFM with the proposed features of repairabilty and reconfigurabilty. Significant power saving is now possible in the multiplier with the introduction of feature of reconfigurability at run time. Self repairability in the multiplier will allow it to recover from logic faults (stuck-at faults) caused by any of 36 4x4 multipliers. The most significant aspect of the proposed architecture is that it will make the FPGAs suitable for performing floating point multiplication operations. The numbers of dedicated 24x24 bit multiplication blocks in the FPGA can be provided according to its suitability for particular DSP operations. The reversible logic design of the CIFM is also proposed as there are a number of advantages associated with reversible logic. The authors believe that proposed work will provide a new direction to FPGAs design both from floating point and reversible logic scenario.

## REFERENCES

[1] GH. A. Aty, Aziza 1. Hussein, I. S. Ashour and M. Mona,"High-speed, Area-Efficient FPGA-Based -Floating-point Multiplier", Proceedings ICM 2003, pp-274-277,Dec. 9-11 2003, Cairo, Egypt.

[2] www.xilinx.com/products/silicon_ solutions/fpgas/virtex/virtex4/

[3] Ahmet Akkas, Michael J. Schulte, "A Quadruple Precision and Dual Double Precision Floating-Point Multiplier",. proceedings DSD 2003,pp.76-81,3-5 September 2003, Belek-Antalya, Turkey.

[4] Himanshu Thapliyal and M.B Srinivas, "Novel Reversible "TSG" Gate and Its Application for Designing Reversible Carry Look Ahead Adder and Other Adder Architectures", Tenth Asia-Pacific Computer Systems Architecture Conference (ACSAC05), Singapore, October 24 - 26, 2005, pp. 805-817.

[5] Himanshu Thapliyal and M.B Srinivas, "Novel Reversible "TSG" Gate and Its Applications for Designing Components of Primitive Reversible/Quantum ALU", Fifth International Conference on Information, Communications and Signal Processing (ICICS 2005), Bangkok, Thailand, 6-9 December 2005,pp.1425-1429 .

[6] Rong Lin and Martin Margala, " Novel Design and Verification of a 16x16-b Self repairable Reconfigurable Inner Product Processor", GLSVLSI'02, April 18-19,2002, Newyork, USA., PP 172-177.

[7] Ronald Scrofano, Gokul Govindu, Viktor Pasanna,"A Library of Parameterizable Floating-Point Cores for FPGAs and Their Application to Scientific Computing", ERSA 2005, Las Vegas, Nevada, USA, June 27-30, 2005,pp.137-148

[8] Gokul Govindu, Viktor K. Prasanna, Vikash Daga, Sridhar Gangadharpalli, V. Sridhar, "Efficient Floating-point Based Block LU Decomposition on FPGAs", ERSA 2005, Las Vegas, Nevada, USA, June 21-24, 2004,pp.137-148

[9] Gokul Govindu, Seonil Choi, Viktor K. Prasanna, Vikash Daga, Sridhar Gangadharpalli, V. Sridhar,"A High-Performance and Energy-Efficient Architecture for Floating-Point Based LU Decomposition on FPGAs", IPDPS 2004, Santa Fe, New Mexico, USA,26-30 April 2004.

[10] Gokul Govindu, Ling Zhuo, Seonil Choi, Viktor K. Prasanna, "Analysis of High-Performance Floating-Point Arithmetic on FPGAs", IPDPS 2004,Santa Fe, New Mexico, USA,26-30 April 2004.

[11] R. Landauer, "Irreversibility and Heat Generation in the Computational Process", IBM Journal of Research and Development, 5, pp. 183-191, 1961.

[12] C.H. Bennett , "Logical Reversibility of Computation", IBM J. Research and Development, pp. 525-532, November 1973.

[13] E. Fredkin, T Toffoli, "Conservative Logic", International Journal of Theor. Physics, 21(1982),pp.219-253.

[14] Md. M. H Azad Khan, "Design of Full-adder With Reversible Gates", International Conference on Computer and Information Technology, Dhaka, Bangladesh, 2002, pp. 515-519.

[15] Hafiz Md. Hasan Babu, Md. Rafiqul Islam, Syed Mostahed Ali Chowdhury and Ahsan Raja Chowdhury ,"Reversible Logic Synthesis for Minimization of Full Adder Circuit", Proceedings of the EuroMicro Symposium on Digital System Design(DSD'03), 3-5 September 2003, Belek- Antalya, Turkey,pp-50-54.

[16] Hafiz Md. Hasan Babu, Md. Rafiqul Islam, Syed Mostahed Ali Chowdhury and Ahsan Raja Chowdhury," Synthesis of Full-Adder Circuit Using Reversible Logic",Proceedings 17th International Conference on VLSI Design (VLSI Design 2004), January 2004, Mumbai, India,pp-757-760.

[17] J.W . Bruce, M.A. Thornton,L. Shivakumariah,P.S. Kokate and X.Li, "Efficient Adder Circuits Based on a Conservative Logic Gate", Proceedings of the IEEE Computer Society Annual Symposium on VLSI(ISVLSI'02),April 2002, Pittsburgh, PA, USA, pp 83-88.